\documentstyle[epsfig,subfigure]{mn}
 at 10truept
\title
     [Intrinsic alignments]
{\vglue-3.0truecm
\centerline{\it\small Accepted for publication in  Monthly Notices}
\vglue 2.5truecm
       Measurement of intrinsic alignments in galaxy ellipticities
\author
     [M.L.~Brown et al.]
     {M.L.~Brown$^1$, A.N.~Taylor$^1$, N.C.~Hambly$^1$ \& S.~Dye$^2$\\
     $^1$ Institute for Astronomy, University of Edinburgh,
     Royal Observatory, Blackford Hill, Edinburgh, U.K.\\
	$^2$Astrophysics Group, Blackett Lab, Imperial College, Prince Consort Road, London, U.K.\\
        mlb@roe.ac.uk, ant@roe.ac.uk}}
\def\bib{\parskip=0pt\par\noindent\hangindent\parindent
    \parskip =2ex plus .5ex minus .1ex}

\newcommand{\be}{\begin{equation}}
\newcommand{\ee}{\end{equation}}
\newcommand{\ba}{\begin{eqnarray}}
\newcommand{\ea}{\end{eqnarray}}
\newcommand{\nn}{\nonumber \\}

\newcommand{\hMpc}{\,h^{-1}{\rm Mpc}}

\newcommand{\sg}{\sigma}
\newcommand{\rgl}{\rangle}
\newcommand{\lgl}{\langle}

\begin{document}

\maketitle

\begin{abstract}
We measure the alignment of galaxy ellipticities in the local Universe over a
range of scales using digitized photographic data from the 
SuperCOSMOS Sky Survey. We find for a
magnitude cut of b$_{\rm J}<20.5$, corresponding to a median galaxy
redshift of $z \approx 0.1$, and $2 \times 10^6$ galaxies, 
that the galaxy ellipticities exhibit a non-zero
correlation over a range of scales between $1$ and $100$
arcminutes. In particular, we measure the variance of mean galaxy
ellipticities, $\sg^2(\theta)$, in square angular cells on the
sky as a function of cell size and find it lies in the range,
$2\times10^{-4} \geq \sg^2(\theta) \geq \sim 1\times10^{-5}$ for cell
side lengths between $15 \leq \theta \leq 100$ arcminutes. Considering the
low median redshift of the galaxies in the sample and hence the
relatively low effective cross-section for lensing of these galaxies
by the large-scale structure of the Universe, we propose that we have
detected an intrinsic alignment of galaxy ellipticities. We compare
our results to recent analytical and numerical predictions made for
the intrinsic galaxy alignment and find good agreement. We discuss the importance
of these results for measuring cosmic shear from upcoming shallow
surveys (e.g. Sloan Digital Sky Survey) and we outline how these measurements could
possibly be used to constrain models of galaxy formation and/or 
measure the mass distribution in the local universe.

\end{abstract}

\section{Introduction}

Over the last decade, considerable interest has been directed towards
the measurement and analysis of galaxy ellipticities as a means to 
estimate shear induced by gravitational lensing (see e.g. Bartelmann \& Schneider 1999). 
On the scale of galaxy 
clusters, this is now a well-established method, with tools to invert
the shear pattern and measure the distribution of Dark Matter. Recently
attention has moved to larger scales to measure the
cosmological weak shear signal of lensing by large-scale structure and 
a number of groups have made consistent detections (Bacon, Refregier \& Ellis 2000;
Kaiser, Wilson \& Luppino 2000; van Waerbeke et al. 2000; Wittman et 
al. 2000). 
Despite the remarkable success of the shear analysis, up until recently very 
little attention had been paid to the prospect of intrinsic alignments
mimicking the gravitational shear signal. The most likely ways for this to 
occur is during the tidally induced spin-up of galaxies (Hoyle 1949), where 
the angular momentum axes, and hence ellipticities, are aligned, or through 
the alignment of galaxy and halo shapes.

Over the last year, this problem has been addressed by a number of groups
using a combination of numerical (Heavens, Refregier \& Heymans 2000, 
Croft \& Metzler 2001) and
analytic methods (Catelan, Kamionkowski \& Blandford 2001, Crittenden
et al 2001).
Although these results are in rough agreement, a complete understanding
of alignments is less secure, with the main problems lying in understanding 
the coupling of the tidal and inertial tensors of dark matter haloes, and the alignment 
of galaxies and haloes. On the 
observational side Pen, Lee \& Seljak (2000) have recently 
claimed a weak detection of spin-spin correlations in the Tully catalogue.
In this letter we have measured the variance of ellipticities of galaxies at
low-redshift, where the effect of intrinsic alignments is predicted to be orders of 
magnitude higher than a lensing effect. These observations may help to distinguish 
between models for alignments.

 This paper is laid out as follows. In Section 2 we present the observational
material used in the analysis. In Section 3 the main analysis methods are
introduced, and our results are presented in Section 4, where we discuss the 
significance of our results. Our conclusions are presented in Section 5.

\section{Observational material}

The observations analysed in this paper are taken from the SuperCOSMOS Sky Survey
program (Hambly et al.~2001a). This digitised photographic sky survey
consists of Schmidt photographic plates
($6^{\circ} \times 6^{\circ}$; plate scale, $1$ cm $= 11.2'$) and
covers the entire southern sky (894 individual Schmidt fields) 
in two colours, b$_{\rm J}$ and~R. The material used in this analysis
consisted of $436$ Schmidt plates corresponding to $\sim 10,000$ square 
degrees.\footnote{Full details and online access
to the data are available via the World Wide Web at URL
{\tt http://www-wfau.roe.ac.uk/sss}}. 

We have constructed an object catalogue (including both stars and galaxies) for
the $436$ fields by pairing the scanned b$_{\rm J}$ and~R plates. Apart from a small
number of large overlap regions near the $0^{\rm h}$ boundary in the survey
which were used for internal consistency checks (see Section 3.4), 
we created a `seamless' catalogue from the overlapping plates using the 
scheme described in Hambly et al.~(2001a). This scheme attempts to include the
image with the best parameters (and exclude the others) when there is a choice
to be made for the same image appearing on more than one plate.  
Pairing the ${\rm b_J}$ and~R plates has the advantage
over single--colour catalogues in that spurious objects on one plate are
eliminated. 
This is particularly important for galaxy studies
from Schmidt photographs since these defects are broken up into many
co-aligned `galaxies' by the image analyser. Such a large source of
contaminants could potentially ruin any shear analysis. 

Regions around bright stars and blended (ie.~multiple)
images have been excluded from the object catalogue to further eliminate
spurious and/or poorly parameterised images. Further details concerning image
parameterisation, classification and photometry are given in Hambly et 
al.~(2001b). Image parameters included in the
final object catalogue generated for this study consisted of (for both
the b$_{\rm J}$ and~R bands) celestial
co-ordinates, local plate co-ordinates, 
second--order moments (semi--major/minor axes and
celestial position angle, see Section 3.1), b$_{\rm J}$ and~R magnitudes,
(b$_{\rm J}$--R) colour,
image classification flag and stellarness index. The external reliability of 
the image classification on the J~plates is demonstrated in Hambly et 
al.~(2001b) as $\geq92$\% reliable for b$_{\rm J}\leq20.5$ with completeness
at around $\sim97$\%. Photometric accuracy for galaxies is around 0.25~mag.
The internal consistency and accuracy in image ellipticity parameters is
demonstrated later in Section 3.4.

\section{Analysis methods}

\subsection{Measuring the ellipticities}

In order to measure the intrinsic alignment of galaxies, we first
divide the survey into square cells of angular side length, $\theta$. We
define the mean ellipticity of galaxies within each cell, $e_i$, as
\ba
	\bar{e}_{1,i} &=& \frac{1}{N} \sum_{j=1}^N e_{1,j}, \hspace{1.cm}
	\bar{e}_{2,i} = \frac{1}{N} \sum_{j=1}^N e_{2,j}, \nn
	e_i^2 &=& \bar{e}_{1,i}^2 + \ \bar{e}_{2,i}^2,
\ea
where $N$ is the number of galaxies in the $i^{th}$ cell and $e_{1,j}$
and $e_{2,j}$ are the ellipticity components of the $j^{th}$ galaxy which
are defined with respect to the cell axes as 
\be
e_{\alpha,j} = \frac{a_j^2-b_j^2}{a_j^2+b_j^2} \left\{ \begin{array}{ll}
\cos 2\varphi_j & \mbox{\ \ }\alpha=1 \\
\sin 2\varphi_j & \mbox{\ \ }\alpha=2 \end{array} \right. \mbox{.}
\ee

Here, $a$ and $b$ are the semi-major and semi-minor
axes of the galaxy and $\varphi$ is the orientation of the semi-major 
axis with respect to the cell axes. 
The weighting scheme used is $w_i=H(I_i-I_{th})$ where $I_i$ is the
measured intensity in pixel $i$, $I_{th}$ is a threshold intensity 
corresponding to the sky background and $H$ is the Heaviside step
function.  
That is, the SuperCOSMOS analyser measures second moments for each
object detected as (Stobie, 1986) 
\ba
Q_{xx} &=& \int d^2\theta \, w_i I(\theta) (\theta_x-\bar{\theta}_x)^2 / \int
d^2\theta \, w_i I(\theta), \nn
Q_{xy} &=& \int d^2\theta \, w_i I(\theta) (\theta_x-\bar{\theta}_x)
(\theta_y-\bar{\theta}_y) / \int d^2\theta \, w_i I(\theta) , \nn
Q_{yy} &=& \int d^2\theta \, w_i I(\theta) (\theta_y-\bar{\theta}_y)^2 / \int
d^2\theta \, w_i I(\theta) ,
\ea
where $\bar{\theta}_x$ and $\bar{\theta}_y$ are the centroids of the
object and are given by
\ba
\bar{\theta}_x &=& \int d^2\theta \, \theta_x  w_i I(\theta) \, / \int d^2\theta
\, w_i I(\theta) , \nn
\bar{\theta}_y &=& \int d^2\theta \, \theta_y  w_i I(\theta) \, / \int d^2\theta
\, w_i I(\theta) \, \, .
\ea
In terms of second moments, the ellipticity components of equation (2)
are 
\ba
e_1 &=& \frac{Q_{xx}-Q_{yy}}{Q_{xx}+Q_{yy}} , \nn
e_2 &=& \frac{2 Q_{xy}}{Q_{xx}+Q_{yy}} .
\ea
The weighting scheme used corresponds to measuring unweighted second
moments within an isophotal threshold for all the objects detected. 
We have varied this threshold limit and have found the resulting
measurements to be largely insensitive to the isophotal
threshold used (see Section 4.2).
To quantify the degree of alignment of the 
cell ellipticities, we calculate the variance of $e_i$ on a given 
scale across the entire survey. Since this statistic has been used to 
measure cosmic shear (e.g. Bacon et al., 2000, Kaiser et al., 2000), 
we can directly compare the contributions of intrinsic and extrinsic 
galaxy alignments.       
In order to obtain an accurate estimate of $e_i$ for a cell, a number
of corrections to the raw image catalogue are required. We now describe 
the sources of error on the ellipticity measurements of galaxies and
the corrections we have applied to the dataset.

\begin{figure}
\vspace{-3.0cm}
\centering
\begin{picture}(200,300)
\includegraphics{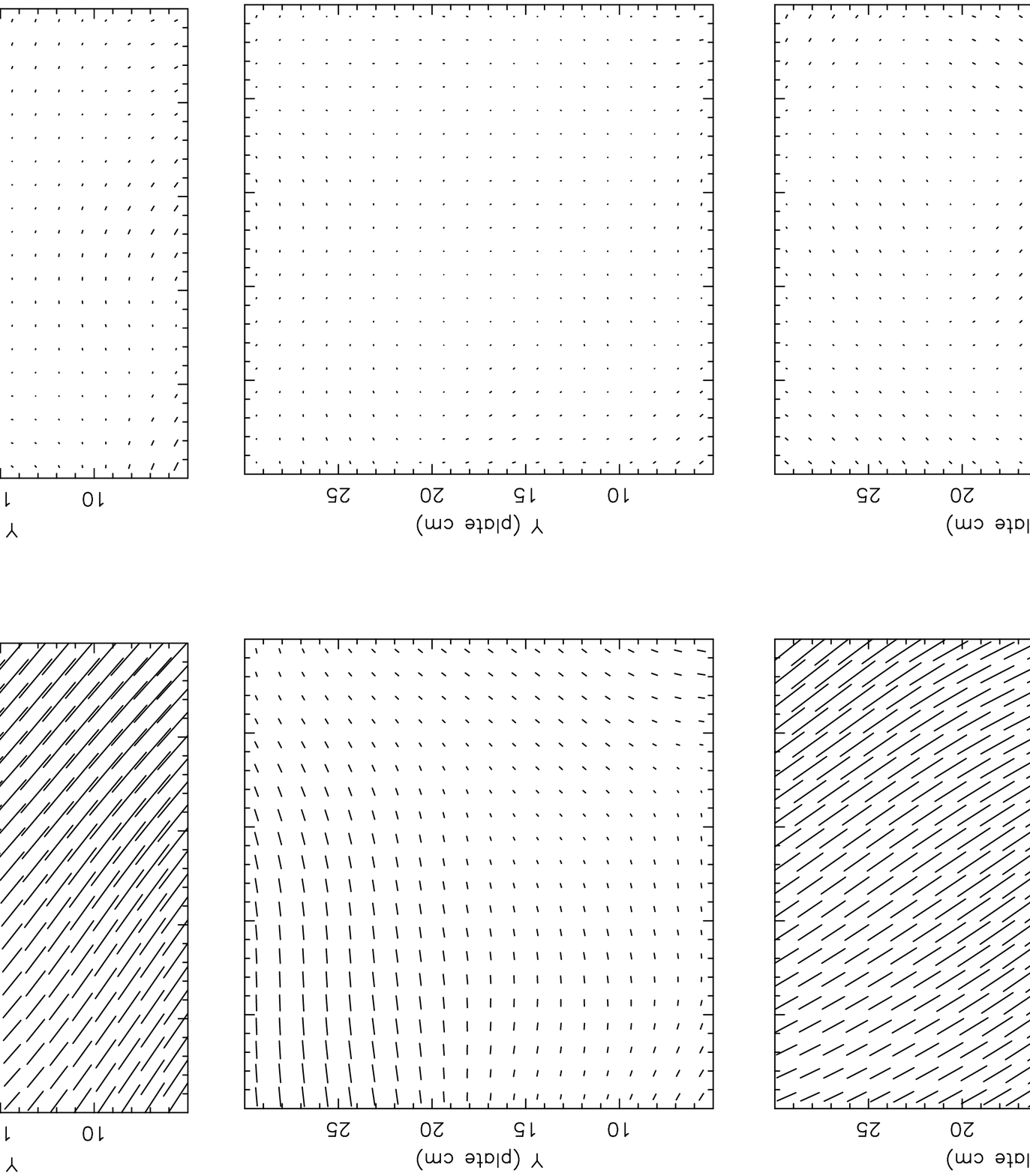}
\end{picture}
\vspace{40mm}
\caption{Ellipticity fields for raw (LHS) and corrected (RHS) stars
for three of the b$_{\rm J}$-band SuperCOSMOS fields. Stars are binned into cells  of
a side $10'$ and smoothed with a Gaussian with a smoothing
scale $15'$. For each plot, the length of each vector drawn is $25$ cm
times the measured cell ellipticity. The average ellipticity in a cell
in the raw fields is $\bar{e} \approx 10^{-2}$, while the average
ellipticity in a corrected cell is $\bar{e}\approx 10^{-4}$. These
plots are typical for the PSF anisotropy distributions on a plate and
the corresponding ~R-band plots are generally very similar.}
\label{fig1}
\end{figure}

\begin{figure}
\vspace{-3.0cm}
\centering
\begin{picture}(200,300)
\includegraphics{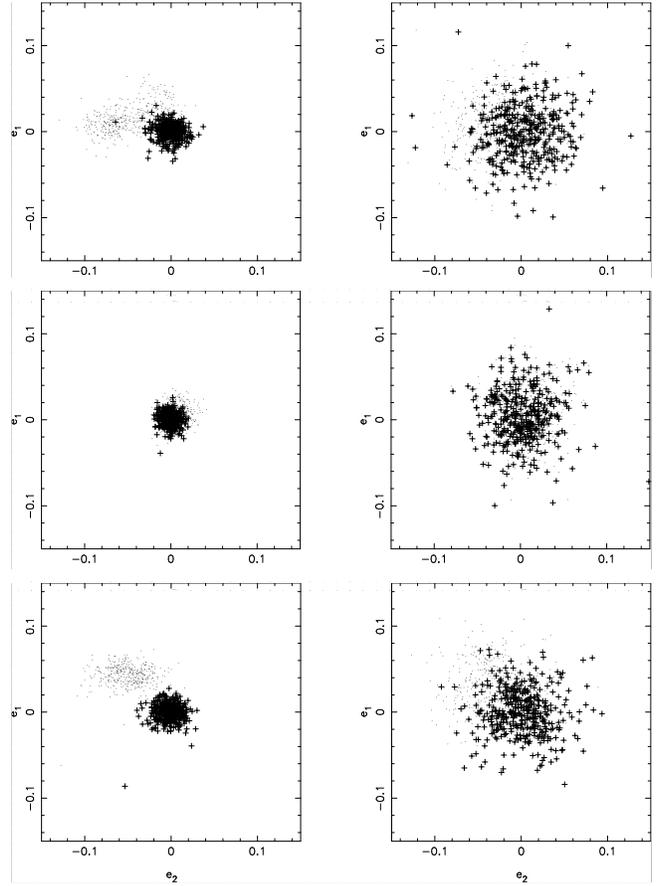}
\end{picture}
\vspace{40mm}
\caption{Ellipticity distribution per cell (stars on the left, galaxies on the
right) for the same three b$_{\rm J}$-band fields shown in figure
1. The dots are the raw stars/galaxies, the crosses the corrected
values. Each cell has an ellipticity estimated
from equation (1) and has an uncertainty given by equation (\ref{error}).
Average stellar distortions in the raw fields of $\bar{e}_1 \sim
\bar{e}_2 \approx 10^{-2}$ are corrected to produce flattened stellar
ellipticity distributions with residual ellipticities, $\bar{e}_1 \sim
\bar{e}_2 \approx 10^{-4}$. Similar ``before'' and ``after'' distributions are
found for the ~R plates.} 
\label{fig2}
\end{figure}

\subsection{Correction for PSF anisotropy}

There are several possible sources of error which could potentially
compromise any shear analysis. These distortions result in spurious 
ellipticities for all the objects detected and must be removed
before a shear analysis can be performed. 
Firstly, there may be slight astrometric distortions present in the
dataset due to emulsion shifts in the photographic plates. However,
these should be negligible --- the dataset we have used has very
precise astrometry and has been used as a standard for making
astrometric corrections in other shear analyses (e.g. Gray et al.,
2001). We have therefore not corrected for astrometric distortions in 
this analysis. For a detailed discussion of the astrometric accuracy of the
catalogue, see Hambly et al., 2001c. 
 
The first correction we have made to the dataset is a correction for
anisotropy in the point spread function (PSF). The are several sources
of error contributing to the PSF anisotropy. Atmospheric dispersion
along with emulsion shifts and mechanical plate distortions, are minor 
contributors to the PSF anisotropy. By far the most important factors, 
however, are tracking errors and field rotation (for an analysis of
the alignment, pointing accuracy and field rotation of the UK Schmidt
Telescope, see Wallace \& Tritton, 1979; further details of the
telescope optics are given in Wynne, 1981). These effects combine 
to produce a systematic PSF anisotropy pattern across each plate which 
needs to be corrected for before the galaxy ellipticity measurements
can be trusted. We have done this by comparing with the ellipticity
field for the stars. Stars should have no intrinsic ellipticity and so 
the measured stellar ellipticities are due to the PSF anisotropy. As 
shown in Kaiser, Squires and Broadhurst, 1995 (hereafter, KSB), the
perturbation to the galaxies' ellipticity components is given by 
\be
\delta e_{\alpha} = P^s_{\alpha \beta} \, p_{\beta} ,
\label{ksb}
\ee  
where $P^s_{\alpha \beta}$ is the ``smear polarizability tensor''
which describes the response of the individual galaxy images to the PSF
anisotropy, $p_{\alpha}$ which in turn can be measured from the stars. 
The analysis described in KSB, which is for a general weighting of
quadrupole moments, is to measure the quantities, $P^s_{\alpha \beta}$
for each individual galaxy and $p_{\alpha}$ from the foreground stars
and to apply the correction in equation (\ref{ksb}) to the individual
galaxies. However, for the essentially unweighted moments that we have
made use of in our analysis (see Section 3.1), $P^s_{\alpha \beta}$
is diagonal with $P^s_{11} = P^s_{22}$ being a measure of the inverse
galaxy size and the KSB correction reduces to 
\be
\delta e_{\alpha} = \frac{Q^*_{xx} + Q^*_{yy}}{Q_{xx} + Q_{yy}} \, 
e^*_{\alpha} ,
\ee
where the $Q^*$'s and the $Q$'s are stellar and galaxy second moments
respectively and $e^*_{\alpha}$ are the measured stellar
ellipticity components. In terms of semi-major and semi-minor axes, this
becomes
\be
\delta e_{\alpha} = \frac{{a^*}^2+{b^*}^2}{a^2 + b^2} \, e^*_{\alpha} ,
\label{ksb2}
\ee
where again, the superscript refers to the stars. 
We have applied the correction in equation (\ref{ksb2}) to individual
galaxy ellipticities using cell-averaged measurements for the ``stellar size''
(${a^*}^2+{b^*}^2$), and where the $e^*_{\alpha}$'s are the average
stellar ellipticity components in the cell. The size of the cells used
for this correction was $10 \times 10$ arcmins. 

\begin{figure}
\centering
\begin{picture}(200,100)
\includegraphics{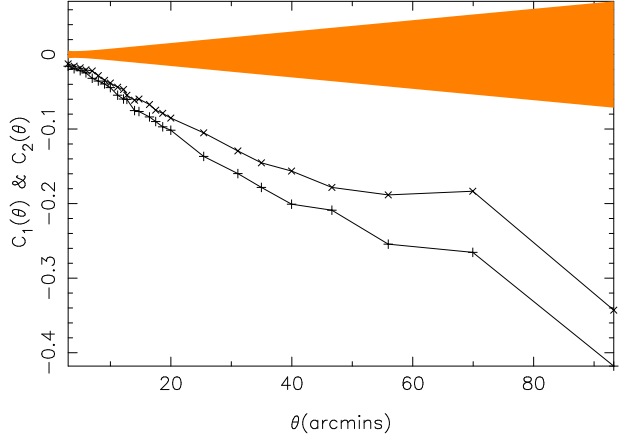}
\end{picture}
\vspace{22mm}
\caption{The linear correlation coefficients, $C_1$ and $C_2$ of the
cell-averaged stellar and galaxy ellipticity components as described in
the text, after correction for PSF anisotropy. The shaded region is the
$3\sigma$ significance level.}
\label{corrfig1}
\end{figure}
  
\begin{figure}
\centering
\begin{picture}(200,100)
\includegraphics{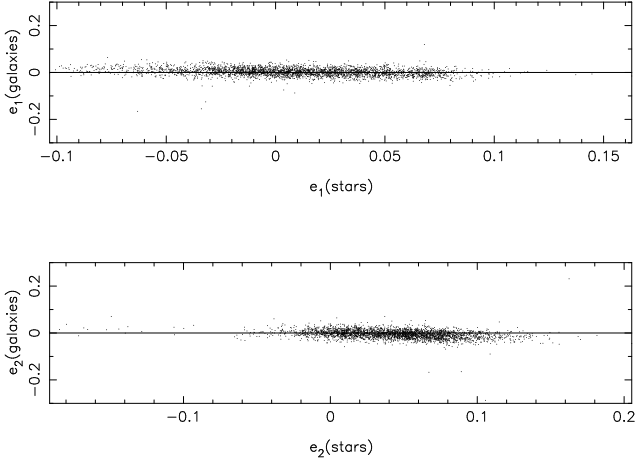}
\end{picture}
\vspace{23mm}
\caption{The observed anti-correlation of corrected galaxy cell
ellipticity components ($e_1$ on top, $e_2$ on bottom) with the
corresponding observed stellar ellipticities.}
\label{corrfig2}
\end{figure}

To perform the correction, the plates are gridded up and an average stellar 
ellipticity is calculated for each cell in the grid. This ellipticity
field is then Gaussian smoothed across the plate. Different smoothing
scales were investigated in order to find an optimum value. The
resulting measurements however did not change significantly for
smoothing scales between $10$ and $30$ arcmins. The individual galaxy
ellipticities are then corrected according to equation (\ref{ksb2}).   
If the grid used for smoothing is too small, a coherent distortion
pattern cannot be made and an essentially random pattern is generated.
Figure 1 illustrates the b$_{\rm J}$ band stellar ellipticities on
three plates before
and after this correction. A strong, coherent distortion of 
$\bar{e}\approx 10^{-2}$ is corrected to produce a
random ellipticity pattern with mean $\bar{e}\approx 10^{-4}$.
Figure 2 shows a scatter plot of the ellipticities ($e_1$ vs. $e_2$)
for both stars (LHS) and galaxies (RHS) before (dots) and 
after (crosses) correction. Each point corresponds to a 
cell, with an ellipticity value given by equation (1), and an
uncertainty given by equation (\ref{error}). 

\vspace{15mm}
\begin{figure}
\centering
\begin{picture}(200,100)
\includegraphics{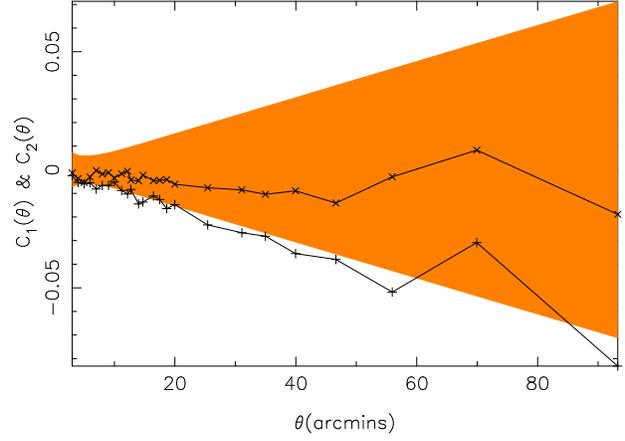}
\end{picture}
\vspace{22mm}
\caption{$C_1$, $C_2$ and the $3\sigma$ significance level after
excluding all galaxies with sizes, $\theta_g < 0.8 \, \lgl \theta_g \rgl$
from the catalogue.}
\label{corrfig3}
\vspace{0.5mm}
\end{figure}

\begin{figure}
\centering
\begin{picture}(200,100)
\includegraphics{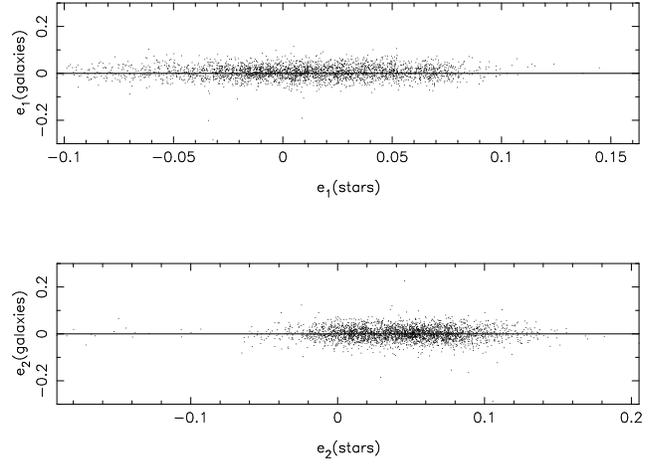}
\end{picture}
\vspace{30mm}
\caption{Cell-averaged galaxy ellipticity components ($e_1$ top, $e_2$
bottom) plotted as a function of the corresponding stellar
ellipticities after applying the galaxy size cut described in the text.}
\label{corrfig4}
\end{figure}

\vspace{-15mm}
To test the success of the PSF correction further we calculated
the linear correlation function of the cell-averaged stellar and galaxy
ellipticity components (c.f. Bacon et al., 2000),
\be
C_i=\frac{\lgl e^*_{\alpha} e_{\alpha} \rgl - \lgl e^*_{\alpha} \rgl 
\lgl e_{\alpha} \rgl} { \sg(e^*_{\alpha}) \sg(e_{\alpha}) },
\label{sgcorr}
\ee
where $e^*_{\alpha}$ and $e_{\alpha}$ are the cell-averaged stellar
and galaxy ellipticity components respectively and $\sg(e^*_{\alpha})$
and $\sg(e_{\alpha})$ are the errors on those two quantities as
measured from the data. In Figure \ref{corrfig1}, we plot $C_1$ and
$C_2$ as measured from the total data set after correction for PSF
anisotropy along with the $3 \sigma$ significance level. It can be
seen from this figure that there is a clear anti-correlation between
the corrected galaxies and the raw stars and furthermore, it is a 
$ \gg 3\sigma$ effect, especially at large bin sizes. This
anti-correlation is clearly seen in Figure \ref{corrfig2} where we
plot the corrected galaxy and raw stellar ellipticity components for a bin size of 
$\theta=70$ arcmins. This anti-correlation is the same effect seen in
the cosmic shear analysis of Bacon et al. (2000) and is due to an
over-correction of the PSF anisotropy for small galaxies. To remove
this effect, we have imposed a cut on galaxy size, only taking
galaxies with $ \theta_g > 0.8 \, \lgl \theta_g \rgl $, where
$\theta_g=a^2+b^2$ is the individual galaxy size, and
$\lgl \theta_g \rgl$ is the mean galaxy size over the whole survey. Only
galaxies satisfying this condition have been used in the final
analysis. The correlation functions, equation (\ref{sgcorr}), calculated after
applying this cut, are shown in Figure \ref{corrfig3}. It is clear from this
figure that the star-galaxy anti-correlation is no longer significant
at the $3 \sigma$ level and this can be seen by eye in Figure \ref{corrfig4}
where the corrected galaxy and raw stellar ellipticity components are
plotted after applying the cut on galaxy size, again
for a bin size of $70$ arcmins.      

\subsection{Seeing correction}

After correcting for PSF anisotropy, we model all other sources of
error (atmospheric turbulence, wind shake etc.) as random
effects. Collectively called seeing, we assume also it is isotropic
and we apply a stochastic correction to remove its effects. 
The effect of seeing  on the galaxy ellipticities is to
circularize the images, causing a decrease in ellipticities. 
The seeing across all of the plates is typically $2''$. This is
comparable to the size of a galaxy near the magnitude limit of our
survey (${\rm b_J} < 20.5$) and thus, needs to be
corrected for before the galaxy ellipticities can be trusted.

We assume that the semi-major and semi-minor axes of each galaxy
transform under the effect of seeing as 
\be
{a'_j}^2 = a^2_j + r_i^2, \hspace{1.cm}  {b'_j}^2 = b^2_j + r_i^2,
\label{seeing}
\ee
where $(a_j,b_j)$ and $(a'_j,b'_j)$ are the axes of the $j^{th}$ galaxy 
before and after the effect of seeing respectively and $r_i$ is the 
average seeing in the $i^{th}$ plate. The average galaxy profile as
measured using the SuperCOSMOS machine deviates from Gaussianity only
very slightly (see Knox et al., 1998) and so equation (\ref{seeing}) 
should describe the effect of seeing on the galaxy images reasonably 
accurately. 

The effect of seeing on the measured ellipticities is
\be
e_{\alpha,j}' = f_{ij} e_{\alpha,j} , 
\label{seecor}
\ee
where $e_{\alpha,j}'$ is the post-seeing ellipticity and  we have defined a
``seeing factor'' which is, in terms of the observed semi-major and
semi-minor axes,
\be
f_{ij} = 1 - \frac{2r_i^2} {a_j'^2+b_j'^2} .
\ee
Here, $a$ and $b$ are the axes of the galaxies after correction for
PSF anisotropy and the uncertainty on individual galaxy sizes was too 
large after this correction to use equation (\ref{seecor})
directly. Instead we estimated an effective galaxy size for all plates
for a given flux cut. We have done this by looking at plate overlap
regions. We have applied the correction in equation (\ref{seecor}) to 
galaxies in these regions and have ``fine tuned'' the effective galaxy 
size so that consistency is achieved in the measurement of both individual
and binned galaxy ellipticities between the overlapping plates (see
Section 3.4). In the case of ${\rm b_J} \le 20.5$, the optimum
effective galaxy size was $2.5''$, which closely matched the directly 
measured mean galaxy size. This was then used to correct all the
plates, using the measured seeing from each plate.

\begin{figure}
\centering
\begin{picture}(200,100)
\includegraphics{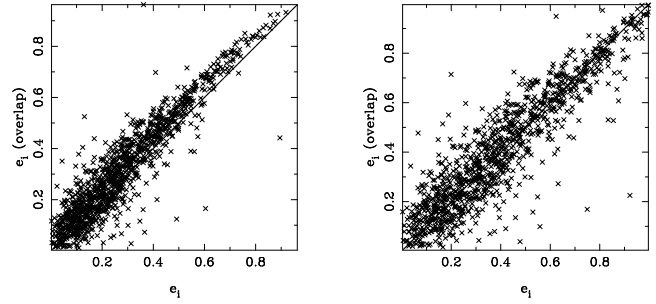}
\end{picture}
\vspace{5mm}
\caption{Individual galaxy ellipticities measured from overlapping
SuperCOSMOS fields. The left-hand panel shows the ellipticity
measurements before the seeing correction is applied. The plate plotted on
the horizontal scale has greater seeing than the plate plotted on the vertical axis.
The right hand
panel shows the ellipticities after the correction.}
\label{olap1}
\end{figure}

\begin{figure}
\centering
\begin{picture}(200,100)
\includegraphics{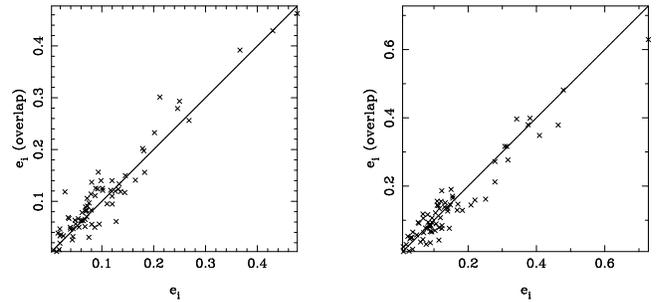}
\end{picture}
\vspace{5mm}
\caption{Binned cell ellipticities, $e_i$ measured from the same
plates as shown in Figure 3. The left-hand and right hand panel are
the measurements before and after the seeing correction is applied
respectively.}

\label{olap2}
\end{figure}

\subsection{Internal consistency tests}

We have tested the seeing correction of the previous section by 
comparing ellipticity measurements in the plate overlap regions. The 
ellipticity measurements for a galaxy in an overlap
region as measured on the two overlapping plates should agree to within
the limits of the measuring process (the measurement errors are dominated
by noise on the original photographs) assuming that we have
corrected for the effect of seeing accurately enough. 

Figures \ref{olap1} and \ref{olap2} show the correlation between
different plates in the overlap region, after correction for PSF
anisotropy, for individual galaxies (Fig. \ref{olap1}) and binned
cells (Fig. \ref{olap2}). The left-hand-side of Figures \ref{olap1}
and \ref{olap2} show the
correlation before correcting for seeing. There is an apparent decrease in the 
ellipticities of the galaxies plotted on the horizontal axis, where the 
seeing on the plate is larger. The right-hand-side of 
Figures \ref{olap1} and \ref{olap2} show the correlation after
correction for seeing. As we are only scaling ellipticities the
scatter is slightly increased, but the correlation is significantly 
greater. These plots are typical for the overlap regions.

\subsection{Comparison with APM Sky Catalogue data}

\begin{figure}
\centering
\begin{picture}(200,100)
\includegraphics{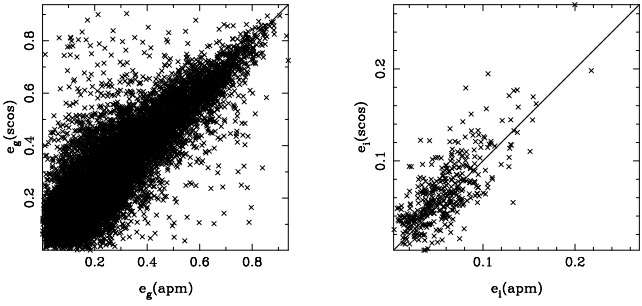}
\end{picture}
\vspace{5mm}
\caption{LHS: Scatter plot of the SuperCOSMOS galaxy ellipticities on a plate against 
those from the APM galaxy survey. RHS: The same as the LHS but for binned cells.}
\label{fig7}
\end{figure}

To check our results further, we have performed a comparison test for
similar alignments in APM data\footnote{available on 
{\tt http://www.ast.cam.ac.uk/$\sim$mike/apmcat/}} in order to test
for any systematic effects introduced by the SuperCOSMOS scanning
procedure. We took a J-R paired star/galaxy catalogue 
from the APM survey for one field and paired it up with the SuperCOSMOS data.
Since the APM data is derived from a copy of the plate that
SuperCOSMOS has scanned, any discrepancies in the scanning procedure
should be highlighted by this test.  
Although the measurement of galaxy shapes for the APM machine is
different, using density measurements rather than the intensity
measurements of SuperCOSMOS, the magnitude cut is low enough that
the difference should not matter.
Figure \ref{fig7} shows a scatter plot for individual galaxy ellipticities for 
a plate on both SuperCOSMOS and APM surveys, and for the binned data, indicating 
a strong agreement between the surveys. 

Having checked for internal and external consistency in the ellipticity catalogue, we
now turn to estimating the variance of the ellipticities in cells of varying
scale, across all of the plates.

\subsection{Estimator for the ellipticity variance}

The variance of the cell ellipticities can be
expressed as the sum of the contributions from all possible sources of
alignment (e.g. Bacon et al., 2000)
\be
\sg_{tot}^2 = \sg_{lens}^2 + \sg_{int}^2 + \sg_{noise}^2 + \sg_{sys}^2
\label{sigvar}
\ee
where we have included the contributions from lensing, intrinsic
alignments, shot noise and systematics. We assume in what follows that the lensing
signal is negligible compared to that from intrinsic alignments for
the median redshift of the galaxies in our sample ($z \approx
0.1$). The
noise term, $\sg_{noise}^2$ is due to intrinsic scatter in galaxy ellipticities and
the random error in the measurement of
the galaxy ellipticities. Since we are averaging over a very large area of sky
($436$ plates covering $\approx  10,000$ sq. degrees with $\approx 1.7 \times 10^6$ galaxies), 
we can expect to beat this term down statistically. 
The final term in equation (\ref{sigvar}) is due to systematic sources
of error. In Section 3.4 we demonstrated the internal consistency of the 
catalogue, indicating that contributions to this term from variations in the catalogue
are small. Further tests are presented in Section \ref{testsys}

We have used a minimum variance estimator for the intrinsic ellipticity
variance, $\sg_{int}^2$ due to intrinsic alignments in excess of the
noise,
\be
	\sg_{int}^2 = \frac{ \sum_i w_i (e_i^2 - {\cal N}_i)}{ \sum_i w_i }
\label{minvarest}
\ee
where ${\cal N}_i$ is the random noise on the estimated cell ellipticity, $e_i$, in 
the $i^{th}$ cell,
and $w_i$ is an arbitrary weighting factor. We assume that the mean
cell ellipticity components are zero (see
Section 4.2). For the moment we will ignore systematic terms.
Following the analysis of van Waerbeke et al. (2000), we estimate the
noise term in the $i^{th}$ cell as 

\be
	{\cal N}_i= \frac{1}{N^2}\left(\sum_{j=1}^{N} e_{1,j}^2 +
	\sum_{j=1}^{N} e_{2,j}^2\right)
\label{error}
\ee
where $N$ is the number of galaxies in the cell.

For a minimum variance estimator, we wish to choose the weights, $w_i$
such that equation (\ref{minvarest}) is minimised with respect to
$w_i$. Denoting the true intrinsic ellipticity variance as 
$\sg_{true}^2$, this optimal weighting scheme is given by
\be
w_i = [\sg^2(\sg_{true}^2) + \sg^2({\cal N}_i)]^{-1} =( 2 \sg_{true}^4 + 2 {\cal N}_i^2 )^{-1},
\label{weight}
\ee 
where $\sg^2(\sg_{true}^2)$ is the uncertainty on the true ellipticity variance,
and $\sg^2({\cal N}_i)$ is the uncertainty on the estimated noise term.
Assuming $\sg_{true}^2$ to be Gaussian distributed, the uncertainty on
the true ellipticity variance will be 
$\sg^2(\sg_{true}^2) = \sg_{true}^4$. The uncertainty in the noise
term can be estimated by noting that the noise contribution from
randomly orientated galaxies will be ${\cal N}_i = e_{rms}^2/N$
where $e_{rms}$ is the root mean squared random ellipticity of the
galaxies. The variance of the noise is then given by
$
\sg^2({\cal N}_i) =  \lgl e_g^4 \rgl / N^2  - e_{rms}^4/N^2
$
where $e_g$ is the ellipticity of the individual galaxies in cell $i$. 
Assuming a Gaussian distribution for the galaxy ellipticities, we
can make the approximation, $ \lgl e_g^4 \rgl \approx 3 \lgl e^2
\rgl^2 = 3 e_{rms}^4 $ and we have 
$
\sg^2({\cal N}_i) = 2 e_{rms}^4/N^2 = 2 {\cal N}_i^2.
$
Combining these variances yields the weighting in equation (\ref{weight}).
Substituting this into the expression for $\sg_{int}^2$ yields the
minimum variance estimator:
\be
	\sg_{int}^2 = \frac{ \sum_i (e_i^2 -{\cal N}_i)/(2 \sg_{true}^4 +2{\cal N}_i^2)}
	{ \sum_i (2 \sg_{true}^4 +2{\cal N}_i^2)^{-1} } .
\label{est}
\ee
The error in this expression is given by 
\be
	\sg^2(\sg_{int}^2) = \frac{1}{\sum_i w_i} = 
	\left( \sum_i ( 2 \sg_{true}^4 + 2{\cal N}^2_i)^{-1} \right)^{-1}.
\label{err}
\ee
Equations (\ref{est}) and (\ref{err}) form the basis of our analysis.

\section{Results}

\begin{figure}
\vspace{15mm}
\centering
\rotatebox{-90}{
\begin{picture}(200,100)
\includegraphics{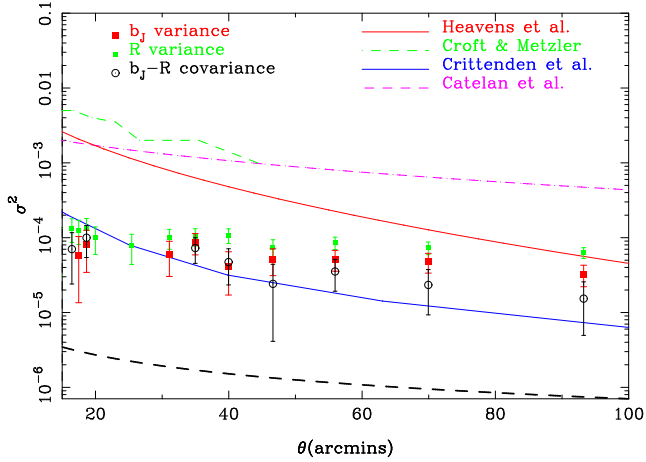}
\end{picture}
}
\vspace{-25mm}
\caption{Measured ellipticity variance, $\sg_{int}^2$ 
over 436 SuperCOSMOS fields as a function of angular scale,
$\theta$ along with the weak lensing prediction (heavy dashed line) for a 
$\Lambda$CDM model and a median redshift equal to that of our galaxy 
sample. Also shown are various predictions for the intrinsic signal - 
see text for details.} 
\label{fig6}
\end{figure}

\subsection{Measurement of the ellipticity variance}

We have used equation (\ref{est}) with $\sg_{true}=0.01$ to
calculate the variance of the intrinsic cell ellipticities,
$\sg^2_{int}$ over a wide range of angular scales. Even so, the
results are fairly insensitive to the value of $\sg_{true}$ used. 
We have measured
both the ${\rm b_J}$ band variance, $\lgl e_i({\rm b_J})e_i({\rm b_J})
\rgl$, and the R band variance, $\lgl e_i(R)e_i(R) \rgl$ as well as
the cross-correlation between the two bands, $\lgl e_i({\rm b_J})e_i(R)
\rgl$. Here, the angled brackets denote the weighted average as in
equation (\ref{est}). Measuring the cross-correlation has the advantage that
systematic effects that are uncorrelated between the two bands are
cancelled out. The results of all these measurements are shown in
Figure \ref{fig6}. In the final analysis, we have made use of
$1.86\times10^6$ galaxies for the ${\rm b_J}$ band variance,
$1.97\times10^6$ for the R band variance and $1.68\times10^6$ for the
covariance measurement.
Note that the scale at which the stellar
ellipticities are smoothed for the PSF anisotropy correction is $15'$
and so, at scales smaller than this, our results may be compromised by 
residual PSF anisotropy distortions. We have, therefore, not plotted
results below this scale in Figure \ref{fig6}. Beyond this scale, we
are confident that our measurements are not dominated by systematics. 
This assertion is supported by the agreement between the two single
band variance measurements and the cross-correlation signal. Further
tests for systematics are presented in Section 4.2. 

\begin{figure}
\vspace{-3mm}
\centering
\begin{picture}(200,100)
\includegraphics{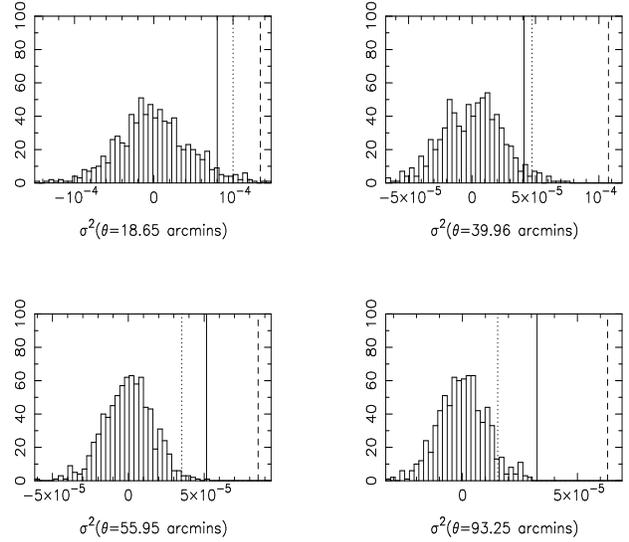}
\end{picture}
\vspace{40mm}
\caption{The variance measurements from the 1000 randomised catalogues
(histograms) along with the measured signal for four different values
of the cell side length, $\theta$. In each case, the full line is the
${\rm b_J}$ band variance, the dashed line is the R band variance and
the dotted line is the covariance measured between the two bands.}
\label{randoms}
\vspace{-0.3cm}
\end{figure}

In addition to using equation (\ref{err}),
we have also estimated the errors on our
measurements from 1000 randomisations of the dataset. In these
randomisations, only galaxies used in the final analysis were
included. The galaxies were assigned a position angle taken from a
random distribution between 0 and 180 degrees while their semi-major
and semi-minor axes were left unchanged. Equation (\ref{est}) was then
used to measure the ellipticity variance from the 1000 randomised
datasets. The resulting distributions of 1000 variance measurements
are centered on zero at all scales, indicating that our estimator is
unbiased. These distributions are shown in Figure \ref{randoms}, along
with the measured signal, for four scales: 19, 40 and 56 and 93 arcmins. 
The errors on the signal measured from the real dataset were
then calculated as the standard deviation of the measurements from the
randomised catalogues. These errors agree quite well with the errors
calculated using equation (\ref{err}) although those taken from
the randomisations are, in general, slightly larger. We have therefore
plotted the $1\sg$ errors from the randomisations in Figure
\ref{fig6}.
     
Our measurements are two orders of magnitude
larger than the signal expected from weak lensing (see e.g. Jain \& 
Seljak, 1997) for a median source redshift of $z \approx 0.1$
corresponding to the magnitude cut of ${\rm b_J} \le 20.5$ which we have 
used in our analysis. In Figure \ref{fig6} we plot our
results along with the predicted weak lensing signal for a cluster normalised
$\Lambda$CDM model, with $\Omega_m=0.3$ and $\Omega_\Lambda =0.7$.
The discrepancy is exhibited over the entire
range of angular scale, $\theta$ suggesting that we have not 
measured extrinsic gravitational lensing. 

\begin{figure}
\vspace{3mm}
\centering
\begin{picture}(200,100)
\includegraphics{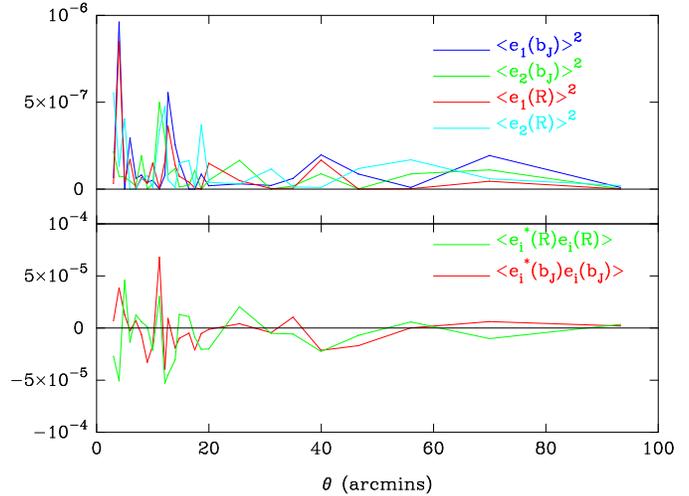}
\end{picture}
\vspace{25mm}
\caption{Top panel: The average galaxy ellipticity components,
$\lgl e_{\alpha} \rgl^2$ over the 436 SuperCOSMOS plates for the ${\rm
b_J}$ and R bands. Bottom panel: The star-galaxy covariance, $\lgl
e_i^* e_i \rgl$ measured
over the whole survey for the two bands.}
\label{means}
\vspace{-0.3cm}
\end{figure}

\subsection{Tests for systematics}
\label{testsys}
\begin{figure}
\centering
\begin{picture}(200,100)
\includegraphics{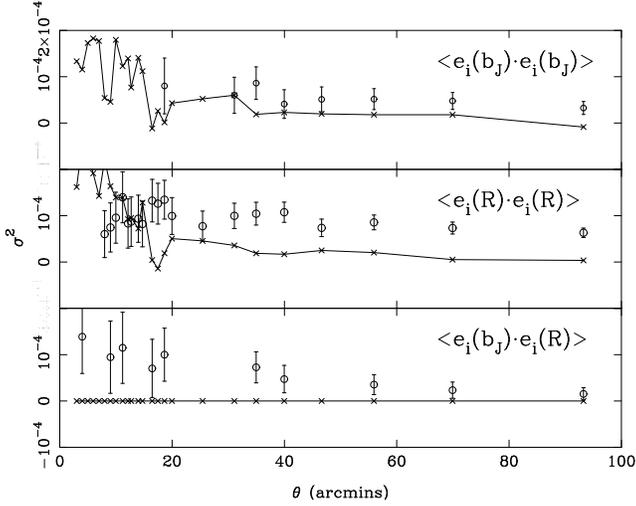}
\end{picture}
\vspace{28mm}
\caption{The signal measured (points with error bars) in ${\rm b_J}$ (top), $R$ (middle) and
the cross-correlation between the two bands (bottom) along with the
cross correlation of $e_1$ with $e_2$ for the galaxies, $\lgl e_1 e_2
\rgl$ (joined crosses). Note the much higher significance of the shear covariance
measurement, $\lgl e_i({\rm b_J}) e_i(R) \rgl$ compared with the shear
variance measurements from the individual bands alone.} 
\label{cross}
\vspace{-3mm}
\end{figure}

To test our results for internal systematics we have estimated
the means of the cell ellipticity components averaged over all
plates. In Figure \ref{means} we 
show that the mean fields are negligible ($\lgl e_\alpha\rgl^2
<10^{-7}$) on all scales.
We have also estimated the star-galaxy covariance, 
\be
	\lgl e e^*\rgl = \lgl e_1 e_1^*\rgl + \lgl e_2 e_2^*\rgl ,
\ee
where $e^*$ is the stellar ellipticity.
This is also shown in Figure \ref{means}, and is well below our results ($|\lgl e e^*\rgl| <10^{-5}$) 
on all scales, indicating that, after applying the cut on galaxy size
(Section 3.2), galaxy ellipticities are no longer correlated with
stellar values.  
We have estimated the cross-correlation of $e_1$ and $e_2$
for the galaxies, $\lgl e_1 e_2\rgl$. This is plotted in Figure
\ref{cross} along with the measured signal for the two single band
variance measurements as well as for the shear covariance signal.  
In all cases, the galaxy cross correlation, $\lgl e_1 e_2 \rgl$ is
consistently below the measured signal. In particular, for the covariance
measurement, we have estimated the galaxy cross correlation as 
\begin{equation}
\lgl e_1 e_2 \rgl=\lgl (e_1({\rm b_J})e_2(R)+e_1(R)e_2({\rm b_J}))/2 \rgl
\end{equation}
and this is consistent with zero at the $10^{-9}$ level on all scales, 
indicating that $\lgl e_i({\rm b_J})e_i(R) \rgl$ is free of significant 
residual systematics and is a robust estimate of the galaxy alignment
signal.

\begin{figure}
\centering
\begin{picture}(200,100)
\includegraphics{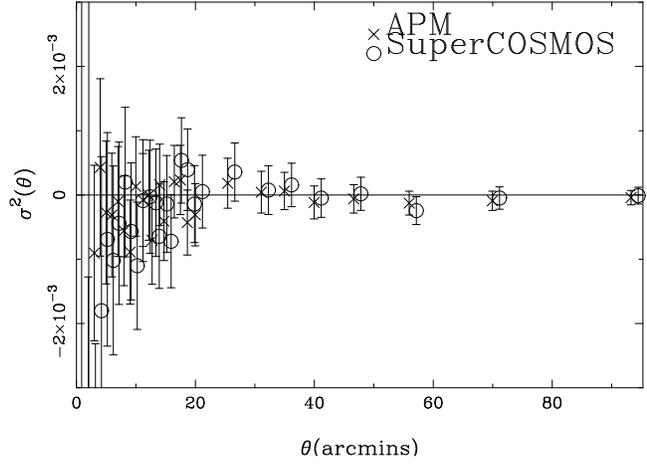}
\end{picture}
\vspace{25mm}
\caption{Comparison of the SuperCOSMOS variance of
ellipticities and the APM variance for one plate (UKST field 78). Note
the errors are much larger than in the final analysis due to the much
smaller number of galaxies measured. The SuperCOSMOS points have been
slightly laterally displaced for clarity.}
\label{apmfig1}
\vspace{-3mm}
\end{figure}

\begin{figure}
\centering
\begin{picture}(200,100)
\includegraphics{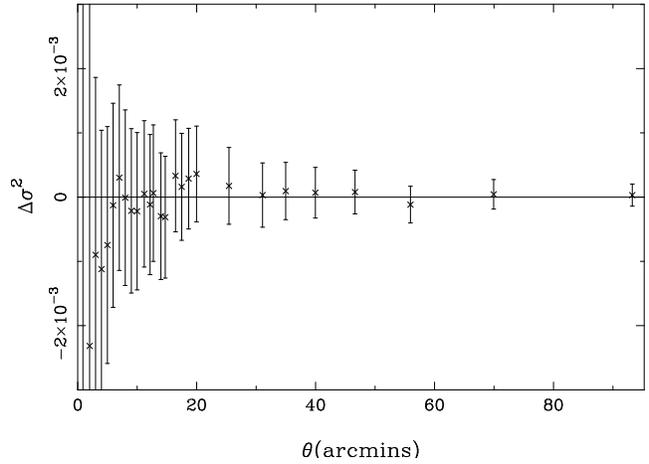}
\end{picture}
\vspace{25mm}
\caption{The difference in the ellipticity variance as
measured from the SuperCOSMOS and APM machines
($\Delta\sg^2=\sg^2_{apm}-\sg^2_{scos}$) for the same field as shown in figure
\ref{apmfig1}. The value of $\Delta\sg^2$ is consistent with zero on
all scales indicating that systematic errors introduced by the plate
scanning procedure are small.}
\label{apmfig2}
\vspace{-3mm}
\end{figure}

As a test for systematics introduced by the measuring process 
we have pushed the APM data (Section 3.5) through our analysis for one plate.
In Figure \ref{apmfig1} we plot the variance as measured by the two
machines for this field. 
Again we find a very good agreement
between both catalogues, suggesting that any systematic effects in our
analysis are small, $\sigma^2_{\rm sys}< 10^{-5}$, and below our
measurement. This agreement between the catalogues is demonstrated in
Figure \ref{apmfig2} where the difference in variance measurements from
the two surveys is plotted. This difference is consistent with zero on
all scales.

\begin{figure}
\centering
\begin{picture}(200,100)
\includegraphics{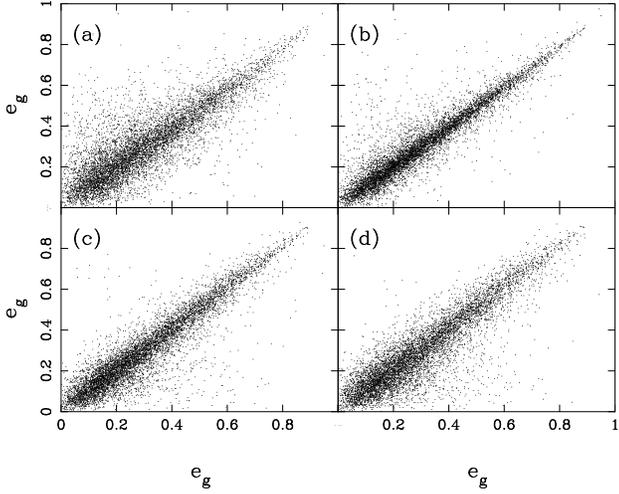}
\end{picture}
\vspace{30mm}
\caption{Individual galaxy ellipticities as measured from UKST field
78 with different isophotal thresholds applied. In each of the four
panels, the horizontal axis measurements are from the Sky Survey data
with an isophotal threshold value of $2.3 \sigma$ above the sky
background. The vertical axes measurements have been made using
isophotal threshold values of (a) $1.5 \sigma$, (b) $2.0 \sigma$, (c)
$3.5 \sigma$ and (d) $4.5 \sigma$ above the sky background.}
\label{thresh}
\end{figure}

As noted in Section 3.1, the SuperCOSMOS machine measures unweighted
second moments of all objects detected within an isophotal threshold,
$I_{th.}$. The nominal value of this threshold in the SuperCOSMOS Sky
Survey is $2.3 \sigma$ above the sky background, $I_{sky}$. To further
establish the reliability of our ellipticity measurements, we have
varied $I_{th.}$ about it's nominal value for one field and compared
the resulting galaxy ellipticity measurements with those from the main
analysis for this field. The results of this test 
are shown in Figure \ref{thresh} showing that the determination of the
galaxy ellipticities is largely insensitive to the isophotal threshold
used to measure the object parameters.

\subsection{Intrinsic alignments}

Recent numerical and analytic studies (Catelan et al., 2001;
Crittenden et al., 2001;  Croft \& Metzler, 2001; Heavens et al.,
2000) have put constraints on the shear signal
expected from intrinsic galaxy alignments. In these studies the
galaxy shape is estimated either from the perpendicular to the
halo angular momentum vector for spirals, or from the shape of the
halo for ellipticals.

In Figure \ref{fig6} we have plotted our results along with the alignment effect 
estimated from each group. Also shown are results from an extension to
the numerical simulations work of Heavens et. al., 2000 for the spiral
galaxy model at $z=0.1$ (A. Heavens, private
communication). The spread in models can be accounted for by the 
choice of mechanism and galaxy type, assumptions about alignments between 
halos and galaxies, correlations between tidal and inertial fields and disc thicknesses.
We have used the redshift scaling $z^{-2n}$, suggested by Crittenden et al., where 
$n$ is the slope of the matter correlation function, to scale the results from 
different redshifts. We have also assumed $\sigma^2(\theta)= 2/(2-n)
C(\theta)$,
where $C(\theta)$ is the ellipticity correlation function.  

Generalising the arguments by Crittenden et al. (2001), at large scales the ellipticity variance
should scale as
\be
	\sigma^2(\theta) \approx A z^{-2n} |1+(\theta/\theta_0)^2|^{-n}.
\ee
One would hope that with future, more precise observations of
intrinsic alignments, it may become possible to constrain the slope of
the matter correlation function, and thus shed light on the
clustering properties of dark matter on these scales ($\le 10 \hMpc$).

\section{Conclusion}

In this paper we have presented a measurement of the intrinsic alignment
effect of galaxies on scales from a few arcminutes to 100 arcminutes. Using
2 million galaxy ellipticities measured from the digitized SuperCOSMOS Sky Survey,
to a depth of ${\rm b_J}=20.5$ and median redshift $z=0.1$ and covering 
436 plates or 10,000 sq. deg. in two passbands, we have corrected the
data for distortions due to PSF anisotropy and seeing, and applied a new
minimum variance estimator to the data. 

After applying these corrections to the dataset and excluding the
over-corrected small galaxies, we find our measurements are
internally consistent, with good agreement in regions where the plates overlap,
with effectively zero star-galaxy ellipticity correlations and zero
$e_1$-$e_2$ cross-correlations. The ellipticity variance measurements from
the ${\rm b_J}$ and R passbands agree well with each other and with
the cross-correlation between the two bands --- further confirmation
that systematic effects are small. We have also demonstrated
external consistency with the APM sky catalogue data over a restricted
region of sky, indicating that we are not contaminated by measurement systematics.

Our resulting estimates of the ellipticity variance over a wide range
of scales are two orders of magnitude higher than that expected from 
gravitational lensing by large-scale structure, but roughly in line with 
those predicted from intrinsic alignments in the gravitational instability 
scenario, although the predictions for intrinsic alignments are still uncertain.
For instance, it is not clear if the effect is dominated by tidal spin or 
shape alignments. This agreement suggests that we are not contaminated
by internal systematics. 

Our results imply that other shallow surveys, such as the Sloan Digital Sky
Survey, should measure roughly the same contribution to the total variance
from intrinsic alignments and  gravitational
lensing shear.  Since the intrinsic alignment signal is expected to scale 
as $z^{-2}$ (Crittenden et al., 2001) and the lensing signal as $z^{1.5}$ 
(Jain \& Seljak, 1997), we expect the ratio of intrinsic alignment to gravitational
shear to scale as
\be
	\frac{\sg^2_{int}}{\sg^2_{lens}} \approx 10^2 \left( \frac{z}{0.1}   \right)^{-3.5}.
\ee
This ratio is unity at around $z\approx 0.37$.
Higher redshift surveys, such as the VISTA Gravitational 
Lensing Survey (Taylor, 2001) with $z \approx 1$, will measure gravitational 
lensing shear.

While one would hope that the correlations between ellipticity, spin and
the shear field would allow one to measure the amplitude of the dark
matter density field, the spin is also determined by the inertial tensor of
the dark matter halo. At present, the relation between shape and the local shear field
of haloes, and the relationship between galaxy ellipticity and halo shape, introduces 
a large uncertainty in our understanding of intrinsic ellipticity alignments. 
With present and future observations and theory we can hope to resolve these issues.

\section*{Acknowledgements}
MLB thanks the University of Edinburgh for a studentship. SD thanks
the PPARC for a PDRA grant, and the Institute for Astronomy, Edinburgh for 
support during the writing of this paper. ANT is a 
PPARC Advanced Fellow. MLB and ANT thank Alan Heavens, Rob Crittenden,
Lance Miller, David Bacon and Alex Refregier for useful
discussions about intrinsic alignments and weak shear
measurements. The authors thank the Institute for Astronomy's Wide
Field Astronomy Unit for assistance in this project.

\section*{References}

\bib Bacon D., Refregier A.,  Ellis R., 2000, MNRAS, 318, 625 
\bib Bartlemann M., Schneider P., 1999, Review for Physical Reports (astro-ph/9909155)

\bib Catelan P., Kamionkowski M., Blandford R.D., 2001, MNRAS, 323,713

\bib Crittenden R., Natarajan P., Pen-U., Theuns T., 2001, ApJ, 559, 552

\bib Croft R.A.C., Metzler C.A., 2001, ApJ, 545, 561
\bib Gray, M.E. et al., 2001, ApJ (accepted)
\bib Heavens A.F., Refregier A., Heymans C., 2000, MNRAS, 319, 649
\bib Hambly N.C., et al., 2001a, MNRAS, 326, 1279

\bib Hambly N.C., Irwin M.J., MacGillivray H.T., 2001b, MNRAS, 326, 1295
\bib Hambly N.C., Irwin M.J., Davenhall, A.C., MacGillivray H.T.,
2001c, MNRAS, 326, 1315

\bib Hoyle, F., 1949, in Burgers J.M., van de Hulst H.C., eds., in Problems
	of Cosmic Aerodynamics, Central Air Documents, Dayton, Ohio, p195

\bib Jain B., Seljak U., 1997, ApJ, 484, 560

\bib Kaiser N., Wilson G., Luppino G., 2000, submitted to ApJL (astro-ph/0003338)
\bib Kaiser N., Squires G., Broadhurst T., 1995, ApJ, 449, 460

\bib Kamionkowski M., Babul, A., Cress, C.M., Refregier, A., 1998,
MNRAS, 301,1064

\bib Knox, R.A.,Hambly, N.C., Hawkins, M.R.S., MacGillivray, H.T.,
1998, MNRAS, 297, 839 

\bib Pen U-L., Lee J., Seljak U., 2000, ApJ, 543, L107

\bib Stobie, R.S., 1986, Pattern Recognition Letters 4, 317
\bib Taylor, A.N., 2001, to appear in Treyer \& Tresse, eds., Where's
the Matter? Tracing Bright and Dark Matter with the New Generation of
Wide Field Surveys, Frontier Group
\bib van Waerbeke L., et al, 2000, A\&A, 358, 30

\bib Wallace, P.T., Tritton, K.P., 1979, MNRAS, 189, 115

\bib Wittman D.M., Tyson J.A, Kirkman D., Dell'Antonio I., Bernstein
G., 2000, Nature, 405, 143

\bib Wynne, C.G., 1981, Q. Jl. RAS, 22, 146

\end{document}